\documentstyle[epsfig,referee]{mn}
\begin{document}
\title{Gravitational Microlensing in NUT Space}
\author [Sohrab Rahvar and Mohammad Nouri-Zonoz]
{ Sohrab Rahvar $^{1,3}$
        and
        Mohammad Nouri-Zonoz $^{2,3}$\\
$^1$ Department of Physics, Sharif University of Technology,\\
 P.O.Box 11365--9161, Tehran, Iran\\
$^2$ Department of Physics, Tehran University, End of North Karegar St,
\\ Tehran 14352, Iran\\
$^3$ Institute for Studies in Theoretical Physics and Mathematics,\\
P.O.Box 19395--5531, Tehran, Iran}
 \maketitle
\begin{abstract}
We study the theoretical signature of magnetic masses on the light
curve of gravitational microlensing effect in NUT space. The
light curves for microlensing events in NUT space are presented
and contrasted with those due to lensing produced by normal
matter. In the next step, associating magnetic mass to massive
astrophysical compact objects (MACHOs), we try to see its effect
on the light curves of microlensing candidates observered by the
MACHO group. Presence or absence of this feature in the observed
microlensing events can shed light on the question of the
existence of magnetic masses in the Universe.
\end{abstract}
\section{Introduction}
Studying the rotational curves of spiral galaxies gives the most
important evidence for the existence of dark matter in the
galactic halo ( Faber \& Gallagher 1979 ; Trimble 1987). Results
from 21 cm band observation shows that for thousands of spiral
galaxies, rotational curves remains constant beyond their luminous
radius (Persic et al. 1996). Comparing luminous matter of universe
$\Omega_{lum} = 0.004$  (Fukugita et al. 1995) with the amount of
baryonic matter $\Omega_B = 0.02 h^{-2} $ (obtained from
nucleosynthesis models of universe) confirms that a major part of
the halo consists of baryonic dark matter ( Copi et al. 1995;
Burles \& Tyler 1998). One of the possible forms of baryonic dark
matter in the halo could be massive astrophysical compact halo
objects (MAHCOs), which are obscure owing to their light mass.
Black holes and Neutron stars can also be considered in this
category. The pioneering idea of using the gravitational
microlensing technique for detection of MACHOs was proposed by
Paczy\'nski \cite{pac86}. Since his proposal, gravitational
microlensing theory entered into its observational phase with
work by several groups. In this paper we study the gravitational
microlensing in an exotic space-time, called NUT space (Newman,
Unti\& Tamburino 1963). The usual gravitational lens effect is
based on the bending of light rays passing a point mass $M$ in
Schwartzschild space-time. In the paper of Nouri-Zonoz \&
Lynden-Bell \cite{nou97} the gravitational lens effect on light
rays passing by a NUT hole has been considered using the fact
that all the geodesics in NUT space including the null ones lie
on cones. It is shown that compared with the Schwartzschild lens,
there is an extra shear due to the gravitomagnetic field that
shears the shape of the source. The effect is shown to be small
even for big values of the magnetic masses (NUT factor). In this
paper we will obtain the gravitational microlensing light curves
in NUT space and compare with the observational light curves of a
few dozen microlensing candidates. The outline of the paper is as
follows. In section \ref{sec:mac}, we give a brief account on the
results of gravitational macrolensing by NUT space and then in
the third section we discuss the microlensing on light rays by
NUT space and, in particular, we find the magnification of a
point-like star. In section \ref{sec:obs} we use observational
microlensing light curves, taken by MACHO collaboration, toward
the Large Magellanic Cloud (LMC) and the Galactic bulge to fit
with theoretical light curves in NUT space. In the final section,
the fitting results are analyzed.
\section{Gravitational Macrolensing in NUT space}
\label{sec:mac} The metric of NUT space is given  (in $t, r,
\theta, \phi$ coordinates) by the line element
\begin{equation}
ds^2=f(r)\left( dt-2l {\rm cos}\theta d\phi\right)^2 - {1\over
f(r)}dr^2 -(r^2+l^2)(d\theta^2 + {\rm sin}\theta d\phi^2)
\end{equation}
where $f(r)=1-{2(Mr+l^2)\over r^2+l^2}$ and $l$ is called the
magnetic mass or NUT factor and one can think of $Q=2l$ as the
strength of the gravitomagnetic monopole represented by the NUT
solution (Lynden-Bell \& Nouri-Zonoz 1998).
% In references [1,2] it is
It was shown that all the geodesics of NUT space, including the
null ones, lie on a cone for which the semi-angle is given by
%\begin{figure*}
%\psfig{file=fig3.eps,angle=-90,width=8.5cm,clip=}
% \caption{ The relation between $r$, $j$ and $L$
% } \label{sn86}
%\end{figure*}

\begin{figure*}
\psfig{file=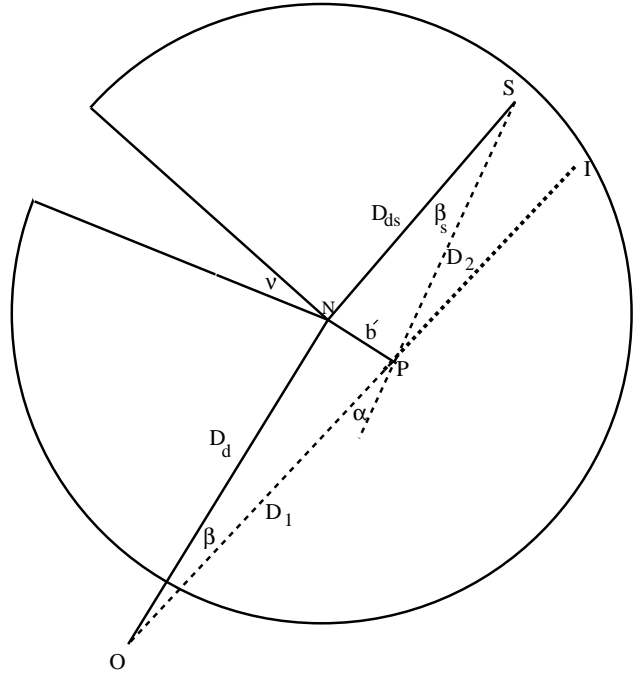,angle=-90,width=8.5cm,clip=}
 \caption{Open, flattened cone and the light ray (dashed line) which is
deflected at $P$ passing the NUT lens. $\nu=2\pi(1-{L\over
(L^2+\varepsilon ^2 Q^2)^{1\over 2}})$ and $\alpha$ are the
deficit and bending angles respectively.
 } \label{sn}
\end{figure*}

\begin{figure*}
\psfig{file=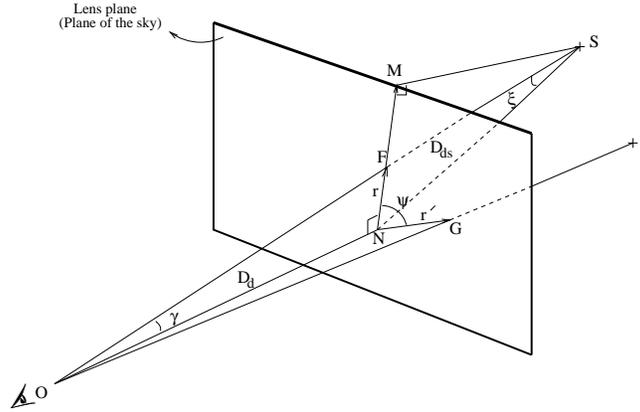,angle=-90,width=8.5cm,clip=}
 \caption{ Lens plane and the position of source $S$, image $I$
 and observe $O$.
 } \label{s}
\end{figure*}
$${\rm sin}\chi = {Q \over b[1+Q^2/b^2]^{1/2}},$$
where $b$ is the impact parameter defined on the cone
(Nouri-Zonoz\& Lynden-Bell 1997). The geometry of lensing in the
case of NUT space could be shown in the following two figures. In
Fig.$1$ the path of a light ray deflected at point $P$ is shown on
an open flattened cone and in Fig.$2$ the positions of the source
and image are shown on the lens plane. The relation between the
positions of source and image is given by
\begin{equation}
\label{rr}
 \frac{r}{r'} = \frac{[4\chi^2 + (\alpha -
\bar\beta)^2]^{1/2}}{\bar\beta},
\end{equation}
where, $r$ and $r'$ denote the positions of source and the image
respectively, $ \bar\beta = \beta (1 + {D_d\over D_{ds}})$ with
the parameters $\beta$, $D_d$ and $D_{ds}$ as defined in Fig. $1$
and $\alpha = 4Gm/bc^2$ is the bending angle defined on the cone.
Using the Jacobian of transformation between image and source
positions, the magnification of the image is given by the
following relation:
\begin{equation}
\label{mag}
 A = {1\over \left[ (1-{\alpha^2
/ {\bar\beta}^2}) (1-{\alpha^2 / {\bar\beta}^2}-8{\chi^2 /
{\bar\beta}^2}) \right]^{1/2}}
\end{equation}
It can easily be seen that for $\chi=0$ one recovers the known
result of the Schwartzschild lens. It is shown that for an
extended source the orientation of the image is also dependent on
the NUT factor through the definition of $\chi$ ( Lynden-Bell \&
Nouri-Zonoz 1997; 1998). Using the above results we study the
microlensing by NUT space in the next section.

\section{Gravitational microlensing in NUT space}
\label{sec:mic} In this section we introduce the basics of
gravitational microlensing by a Schwartzschild lens and then study
the same effect in NUT space.
\subsection{ Basics of gravitational microlensing}
Considerable gravitational lensing occur when the impact parameter
of the light rays is small enough. Since in gravitational
microlensing the deflection angle is too small, for present
telescopes it is impossible to resolve the two images produced
and its effect is only on the magnification of background star.
This magnification is given by:
\begin{equation}
\label{pac}
A(t) = \frac{u(t)^2 +2}{u(t)\sqrt{u(t)^2 + 4}},
\end{equation}
where $u(t) = \sqrt{u_0^2 + (\frac{t - t_0}{t_E})^2}$ is the
impact parameter (position of the source in deflector plane
normalized by Einstein radius) and in which $t_E$ is the Einstein
crossing time (duration of event) defined by $t_E = R_E/v_t$,
where $v_t$ is the transverse velocity of deflector with respect
to the line of sight. The Einstein radius is given by $ R_E^2 =
\frac{4GMD}{c^2}$, where M is the mass of the deflector and $D =
\frac{D_{l}D_{ls}}{D_{s}}$. It is seen that the light curve is
symmetrical with respect to time and since gravitational lensing
effect is independent of the frequency of light, we would expect
the same magnification throughout the spectrum. The probability
of observing a microlensing event is very low (e.g. toward the
Large and Small Magellanic Clouds is about $10^{-7}$) and the
rate of microlensing events also depends on the galactic models
(Rahvar 2003). Comparing the rate of observed microlensing with
what have been expected from theoretical galactic halo models
reveal that only 20 per cent of the halo is made of MACHOs (
Lasserre et al. 2000; Alcock et al. 2000).
\subsection{Gravitational Microlensing in NUT space}
\begin{figure*}
\psfig{file=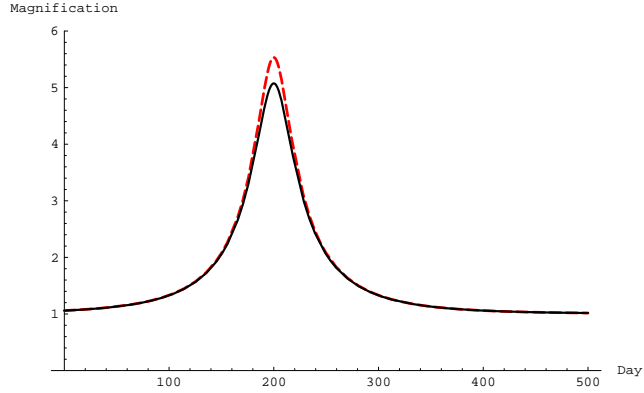,angle=0,width=8.5cm,clip=}
 \caption{ Solid line represents the gravitational microlensing in Schwartzschild metric
 with the parameters of  $u_0 = 0.2$, $t_E = 100$ and $t_0 = 200$
 and dashed line shows the light curve in NUT space with
 $R = 0.2$, $u_0 = 0.2$, $t_E = 100$ and $t_0 = 200$ parameters.
 The magnification of light in NUT space is more that Schwartzschild
 space with the same impact parameter.
 } \label{nut1}
\end{figure*}

In the Galactic scales the configurations of gravitational lensing
have dynamical behavior and this makes gravitational microlensing
light curves very sensitive to the parameters of the space-time
under consideration. In what follows, we find the magnification
function for gravitational microlensing in NUT metric and compare
it with Schwartzschild microlensing. Using Eq. (\ref{mag}) for the
magnification of a point like source in NUT space, it can be
written in the following form:
\begin{equation}
\label{a1}
 A(u_i) = \left[(1-\frac{1}{u_i^4})(1-\frac{1}{u_i^4} -
\frac{8R^4}{u_i^4})\right]^{-\frac12}
\end{equation}
where $R =\frac{R_{NUT}}{R_E}= c\sqrt{\frac{l}{2GM}}$ and $u_i$
indicates the position of the image in the lens plane (normalized
to Einstein radius). In which we define the NUT radius to be
\begin{equation}
R_{NUT}^2 = 2 l D.
\end{equation}
Here we are interested in obtaining the magnification of the
background star as a function of the position of the source in the
lens plane in the absence of the lens. Using definitions of the
Einstein and NUT radii and normalizing all the length scales to
Einstein radius, Eq.(\ref{rr}) can be written in the following
form:
\begin{equation}
\label{u1}
u_s^2 = 4\frac{R^4}{u_i^2} + (\frac1u_i- u_i)^2,
\end{equation}
where, $u_s$ and $u_i$ are the positions of the source and the
image on the lens plane respectively. Hereafter, we omit the
index s of $u_s$ for convenience. In principle, Equation
(\ref{u1}) has the following two solutions:
\begin{equation}
\label{uu}
 \frac{1}{{u_i^{\pm}}^2(u)} = \frac{ 1 +
u^2/2\pm\sqrt{(1+u^2/2)^2 - (4R^4+1)}}{4R^4+1},
\end{equation}
corresponding to the positions of the two images produced by the
lens provided $u^2>2(\sqrt{1+4R^4} - 1)$. Now using Eq.(\ref{a1})
the magnification for each of the images can be written in the
following form:
\begin{equation} \label{aa}
 A^{\pm} =
\frac{(1-\frac{8R^4}{{u_i^\pm}^4-1})^{-1/2}}{|1-\frac{1}{{{u_i}^\pm}^4}|},
\end{equation}
 Substituting equation (\ref{uu}) into equation
(\ref{aa}), the total magnification is:
\begin{eqnarray}
\label{magnif}
 A(u) &=& |A^-| + |A^+| \\
      &=&
\frac{1}{(1-\frac{{{(2+{{{u}}^2}-{\sqrt{-16{R^4}+4
{{{u}}^2}+{{{u}}^4}}})}^2}}{4 {{(1+4 {R^4})}^2}}) {\sqrt{1-\frac{8
{R^4}}{-1+\frac{4 {{(1+4 {R^4})}^2}}{{{(2+{{{u}}^2}-{\sqrt{-16
{R^4}+4 {{{u}}^2}+{{{u}}^4}}})}^2}}}}}} \nonumber  \\ &-&
\frac{1}{(1-\frac{{{(2+{{{u}}^2}+{\sqrt{-16 {R^4}+4
{{{u}}^2}+{{{u}}^4}}})}^2}}{4 {{(1+4
{R^4})}^2}}){\sqrt{1-\frac{8{R^4}}{-1+\frac{4 {{(1+4
{R^4})}^2}}{{{(2+{{{u}}^2}+{\sqrt{-16 {R^4}+4
{{{u}}^2}+{{{u}}^4}}})}^2}}}}} }, \nonumber
\end{eqnarray}
where we use the fact that $A^+$ is negative for
$u^2>2(\sqrt{1+4R^4} - 1)$ . For $R=0$ one can recover
Paczy\'nski's relation, Eq. \ref{pac}. Expanding equation
(\ref{magnif}) in terms of $R^4$ we obtain the following simple
expression for the magnification:
\begin{equation}
A(u) = \frac{2 + u^2}{u\sqrt{4 + u^2}} +
\frac{8R^4(2+u^2)}{u^3(4+u^2)^{3/2}}+{\mathcal{O}}(R^8)+ ...
\end{equation}
In Fig. \ref{nut1} the gravitational microlensing light curves in
NUT and Schwartzschild spaces are shown. In the next section we
use realistic light curves of microlensing candidates towards the
Large Magellanic Cloud and the Galactic bulge to test their
compatibility with theoretical light curves in NUT space.
\section{Compatibility of Microlensing in NUT Space with Observations}
\label{sec:obs} 
\begin{figure}
\psfig{file=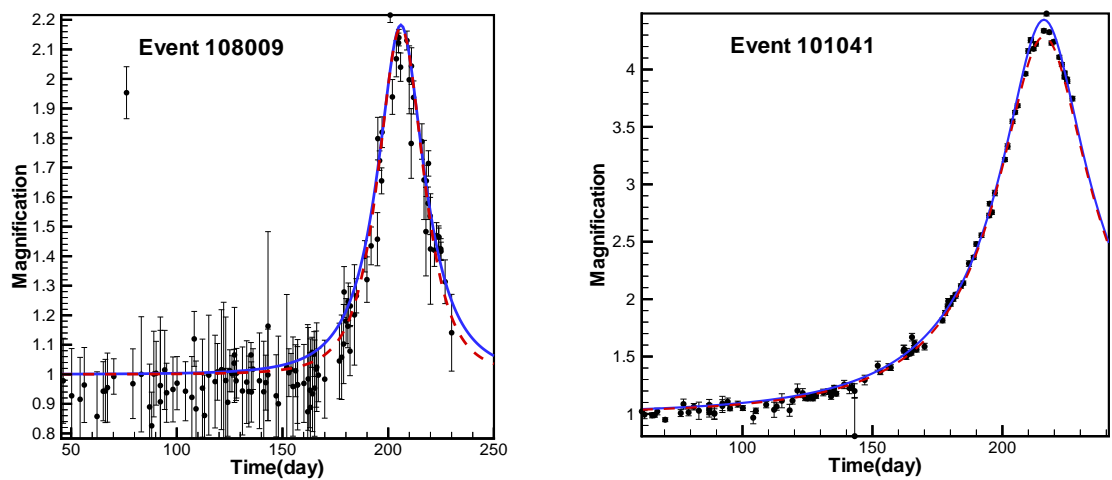,angle=0,width=16.cm,clip=}
 \caption{
Left and right panels show the observed light curves of events
(108009) and (101041) by MACHO collaboration. The best fit of
gravitational microlensing light curves are indicated in NUT
(dashed line) and Schwartzschild (solid line) spaces.}
\label{obs_light}
\end{figure}

\begin{figure}
\begin{center}
\psfig{file=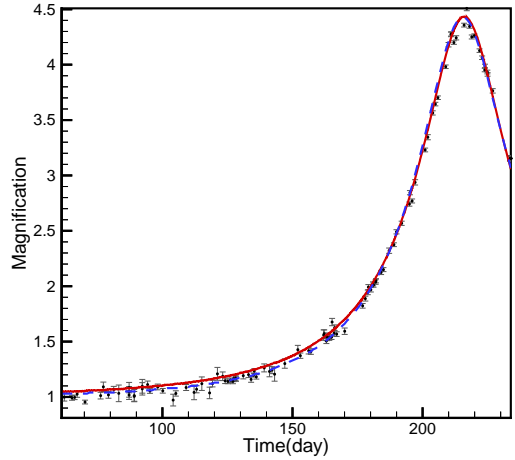,angle=0,width=8.cm,clip=}
 \caption{ Dashed-line and solid line indicate the best fit to
the light curve of event (101041) using the parallax effect and
standard microlensing respectively. The reconstructed parameters
including the parallax effect are obtained to be: $u_0 = 0.36$,
$t_0 = 221 day$, $t_E = 64$, $\widetilde{v} = 0.08 A.U/day$ and
$\theta = -0.65 Rad$. } \label{par_light}
\end{center}
\end{figure}

In this section we compare the light curves of
microlensing candidates, observed by MACHO collaboration, with
those obtained (in the previous section) from our theoretical
study of light curves in NUT space. 44 microlensing light curves
towards the Galactic bulge \cite{alc97} and LMC \cite{alc97b} have
been analysed. These data have been obtained from MACHO group's
database available on the
net\footnote{http://wwwmacho.mcmaster.ca/}. In order to increase
the sensitivity of fitting to the light curves, we express the
light curves in terms of the magnification rather than magnitude
of the background star. Tables (\ref{lmc_nut}) and
(\ref{bulge_nut}) show the results of our analysis of fitting data
with the light curves in the Schwartzschild and NUT metrics. In
the Table (\ref{lmc_nut}), it is seen that including the NUT
charge do not improve the fitting parameter $\chi^2/N_{d.o.f}$,
but in some cases like lmc1b, lmc7 and lmc8 we obtain a non-zero
value for the $R$. This can be interpreted through the degeneracy
problem that arises in the fitting. For two event in the Galactic
bulge candidates, fitting is improved by the inclusion of the NUT
factor. In the event $(101041)$, $\chi^2$ is about
$453/105_{d.o.f}$ from NUT fitting, while in the Schwartzschild
space $\chi^2$ is $570/106_{d.o.f}$. The reconstructed parameter
of magnetic mass from this fit is $R = 0.41$ with an uncertainty
of $0.031$ from covariance matrix. In the second event $(108009)$
the goodness of fitting to NUT is weaker than the event
$(101041)$ and the value of $\chi^2$ is $278/118_{d.o.f}$ in the
NUT space compared to $282/119_{d.o.f}$ in the Schwartzschild
metric. Fig. ({\ref{obs_light}}) shows the light curves of the
microlensing candidates (101041) and (108009) with the best
fitting in NUT and the Schwartzschild spaces.
%Tables. (\ref{lmc_nut}) and (\ref{bulge_nut}) also represent the
%best $\chi^2$ and the reconstructed parameters from the fitting
to the LMC and galactic bulge stars respectively.\\
%It is easily seen that the only difference in the fitted light
%curves is the higher peak in the case of NUT microlensing and for
%the large impact parameters the two light curves are coincident.
To test the reliability of NUT fitting to the light curve of event
(101041), we also tried to fit this light curve with another well
known effect, the so-called non-standard microlensing light curve
in the Schwartzschild metric. Since the duration of this event is
long enough, the parallax effect should be taken into account. If
the variation of the velocity rotating component of the Earth
around the sun is not negligible with respect to the projected
transverse speed of the deflector, then the apparent trajectory
of the deflector with respect to the line of sight is a cycloid
instead of a straight line. The resulting amplification versus
time curve is therefore affected by this parallax effect (
Grieger {\it et al} 1986; Gould 1992).
% It is seen that the using NUT metric impove the value of $\chi^2$, but
%can this light curve be fitted with so-called  like considering
%parallax or finite
%size effect?\\
Analysing this event, taking into account the parallax effect,
shows the goodness of fit $\chi^2 = 356/104_{d.o.f}$ which is
compared to the standard microlensing in Fig.(\ref{par_light}).
Here we obtain the transverse speed of deflector on the ecliptic
plane $ \widetilde{v} = \frac{v}{1-x} = 0.08$ $ au$ $ d^{-1}$. It
is seen that the parallax effect improves the fit more than the
inclusion of the magnetic mass (NUT factor).\\

% Table.
%\ref{micro_nut} shows that some of the events like lmc 1b, lmc 5,
%lmc 6 and lmc 8 have been fitted reasonably well with the NUT
%space light curve, indeed they are better than the similar
%results given in Table. \ref{micro_sch} for Schwarzschild light
%curves. For these four events $R$ have been found in the domain
%of $R\in[0.01,0.68]$ witch corresponds to the magnetic mass domain
%$l\in\frac{2GM}{c^2}[10^{-4}, 4.6\times10^{-1}]$. Using $0.2$
%solar mass for the mean mass of MACHOs from microlensing
%experiments, one can obtain the domain $l\in[10,2500]cm$.

\begin{table*}
\caption[]{ light curve of microlensing candidates towards LMC
have been fitted with microlensing in Schwartzschild and NUT
space. The result of fit is shown with the best $\chi^2$ and
reconstructed parameters}
\begin{center}
\begin{tabular}{lclclclclclclclclclclcl}
\hline
%\noalign{\smallskip}
$Event :$  &${\chi^2}^{NUT}$& $N_{d.o.f}^{NUT}$ & $u_0^{NUT}$ & $t_0^{NUT}$ & $t_E^{NUT}$ &$R^{NUT}$ &${\chi^2}^{Sch}$ & $N_{d.o.f}^{Sch}$&$u_0^{Sch}$ & $t_0^{Sch}$ & $t_E^{Sch}$  \\
%$Event :$  & $u_0^{Sch}$ & $t_0^{Sch}$ & $t_E^{Sch}$ & $\chi^2/N_{dof}^{Sch}$ & $u_0^{NUT}$ & $t_0^{NUT}$ & $t_E^{NUT}$ & $R^{NUT}$ &$\chi^2/N_{dof}^{NUT}$ \\
\hline \hline
lmc 1a   & 1841 &495 & 0.13 & 56  & 17 & 0.0044& 1845& 496& 0.13 & 56  & 17 \\
lmc 1b   &412   &389 & 0.12 & 57  & 16 & 0.3   & 421 &390 & 0.12 & 57  & 17  \\
lmc 4    &861   &261 & 0.34 & 646 & 19 & 0.031 & 864 & 262& 0.34 & 646 & 19   \\
lmc 5    &209   &264 & 0.025& 24  & 27 & 0.0004& 206 & 265&0.025 & 25 & 27    \\
lmc 6    &321   &397 & 0.44 & 197 & 50  & 0.01 & 322 & 398&0.44  & 197 & 50  \\
lmc 7    &832   &266 & 0.67 & 463 & 27 & 0.81  & 838 & 267& 0.2   & 463 & 51   \\
lmc 8    &315   &261 & 0.89 & 389 & 27 & 0.68  & 317 & 262& 0.51 & 389 & 34  \\
\noalign{\smallskip} \hline
\end{tabular}
\end{center}
\label{lmc_nut}
\end{table*}

\begin{table*}
\caption[]{ light curve of microlensing candidates towards
Galactic Bulge have been fitted with microlensing in
Schwartzschild and NUT space. The result of fit is shown with the
best $\chi^2$ and reconstructed parameters}
\begin{center}
\begin{tabular}{lclclclclclclclclclclcl}
\hline
%\noalign{\smallskip}
$Event :$  &${\chi^2}^{NUT}$& $N_{d.o.f}^{NUT}$ & $u_0^{NUT}$ & $t_0^{NUT}$ & $t_E^{NUT}$ &$R^{NUT}$ &${\chi^2}^{Sch}$ & $N_{d.o.f}^{Sch}$&$u_0^{Sch}$ & $t_0^{Sch}$ & $t_E^{Sch}$  \\
\hline \hline
101001     & 1143 & 103  & 0.08 & 160 & 24   & 0.00044 & 339 &104  & 0.12  & 161  & 26.7 \\
101041     & 453  & 105  & 0.35 & 216 & 64   & 0.41    & 570 & 106 &0.23  & 216  & 71    \\
101044     &554   & 104  &0.18  & 203 & 14   & 0.     &  547&105  & 0.18 & 203   &14 \\
101046     &112   & 107  &0.017 & 177 & 6.22 & 0. &112 &108  & 0.01   &177     &6.23 \\
104013     &154   & 110  &1.65  & 161 & 4.23 & 0.98   &154 &111 &0.77 &161 &6.8 \\
104036     &58    &  96  &0.31  &  103 & 13  & 0.      & 58 & 97 & 0.31& 103 &13.5\\
104037     &4510  &  104 & 0.18 &  116 & 89  & 0.26    & 4692& 105 &0.14 &116& 92\\
108009     &275   & 118   & 1.67 &  206 & 10  & 1.23    & 282  &119 &0.5& 206& 23\\
108024     &134   & 108   & 0.35 &   203& 20  & 0.0003  &  116 &109 &0.36 &203 &20\\
108054     & 1621 & 111   & 0.05 &   196& 9.7 & 0.031   &1621  &112 & 0.05&196& 9.7\\
110003     & 91   &80     &0.44  & 94    &4.89&   0.0054&91    &81 &0.44 &94 &4.8\\
110008     & 50   &77     & 0.59 & 107   & 5.6& 0.0044  &50    & 78 &0.6 &107 &5.6\\
110011     & 74   &60     &0.34  & 129   &9.7 &0.00008  & 74 & 61& 0.34 &129 &9.7\\
110055     &44    &76     &0.49  &  166  &10.8& 0.031   &44  &77& 0.49 &166 &10.8\\
111029     & 52   & 53    &0.42  &  177  & 5.9& 0.07    & 52 & 54& 0.42 &177& 6\\
111039     & 34   & 55    & 0.4   & 254 & 58  & 0.054   &34  &56& 0.40 &254& 58\\
113026     &  91  & 80    & 0.44  &  94 & 5   & 0.      &  91 & 81& 0.44& 94 &4.8\\
113014     & 310  &110    &0.67   & 161 & 7.6 & 0.      &154 & 111& 0.77& 161& 6.8\\
113023     &139   &146    &0.32    &188 &8.8   &0       &  139& 147& 0.32& 188& 8.8\\
114021     &  111 &116    &0.83    &168  &27   &0        &  111 &117 &0.83 &168& 27\\
114042     & 223  & 110   &0.27    &181 &8.27  &0        &223 &111& 0.27 &181& 8.27\\
115026     &201   &64     &0.38    &87  &16    & 0.1      &201&65& 0.38& 87& 16\\
118026     &102   &117    &0.38    &77  &11.8   &0.031   &102 &118& 0.28& 77 &11.8\\
118038     &100   &118    &0.49    &80  &24      &0.04   &100 &119& 0.49& 80& 24\\
119001     & double  Lens \\
119005     &90    &118    &0.46    &215 &6.8    &0.01    &90 &119& 0.46& 215& 6.8\\
119053     &133   &105    &3.45    &156 &4.63    &2.21   &136 &106& 0.53 &156& 17\\
120007     &28    &71     & 2.41   & 161& 2.97   & 2.12  & 33 &72&0.25& 161 &14.3\\
121026     &206   &71     &0.18    & 94 & 13     &0.028   &206  &72 &0.18 &94 &13\\
124002     &248   &63     &0.00004 & 8.3& 40     &1.24    &226  &64& 0.02& 42& 68\\
124031     &72    &64     &0.02    &168 &16      &0.17    &34 &65 &0.000002 &167 &15\\
128027     &234   &96     &0.14    &221 &8.34    &0   &234 &97& 0.14 &221 &8.34\\
128055     &328   &98     &0.49    &143 &15      &0.044   &328 &99& 0.49& 143& 15\\
159053     &362   &40     &0.17    &130 &25      & 0.26    &367 &41& 0.13& 130& 26\\
162006     &283   &46     &0.36    &145 &11      &0.044    &283& 47 &0.36& 145& 11\\
167004     &73    &45     &0.33    &108 &18      &0        &72 & 64 &0.33 &108 &18\\
 \noalign{\smallskip} \hline
\end{tabular}
\end{center}
\label{bulge_nut}
\end{table*}
\section{Summary}
\label{sec:conc} In this article we have studied the gravitational
microlensing in NUT space with the aim of learning more concerning
the magnetic masses and their observability regarding MACHOs. In
the first step we introduced the ratio $R = \frac{R_{NUT}}{R_{E}}$
which in a sense is the ratio of magnetic mass of the lens to its
mass. Then we found the magnification for the microlensing in NUT
space in terms of this parameter. In the next step comparing the
light curves in NUT space with a set of observational light
curves from MACHO collaboration. We showed that the inclusion of
magnetic masses in the form of the NUT metric as the surrounding
space--time of the lens, instead of the usual Schwartzschild lens
in most cases will not change the fit. Even in those cases where
the fit is changed one finds that the fit would be much better
with the inclusion of the parallax effect.\\
In the above considerations we have assumed that there are
magnetic mass constituents of MACHOs and studied their
observability through microlensing effect. In a recent paper
(Bradley et al 1999) global solutions composed of locally perfect
fluids that are matched with the NUT metric and interpreted as it
source are presented. Owing to the fact that mass of MACHOs could
range from that of a large planet to a few $M_{\odot}$, we do not
see any argument against the conjecture that the fluid cores
discussed in that paper could be probable candidates for MACHOs. \\
Although the above consideration shows that the effect of NUT
factor is almost negligible but one can not rule out the existence
of NUT charge on that basis. As a matter of fact with the next
generation of microlensing experiments in which both the
photometric precision and sampling will be improved, the
(non)-existence of NUT charge could be addressed with much more
precision. A Monte-Carlo simulation for observability of magnetic
masses in the next generation microlensing experiments is in
progress.\\

ACKNOWLEDGMENT\\
The authors thank Marc Moniez for his useful comments on the
analysis of microlensing light curves.

\end{document}